\documentclass[reprint,aps,prl,twocolumn,superscriptaddress,preprintnumbers,letterpaper,natbib,floatfix]{revtex4-1}
\pdfoutput = 1

\usepackage{amssymb}
\usepackage{amsmath}
\usepackage{epsfig}
\usepackage{comment}
\usepackage{color}
\usepackage{hyperref}
\usepackage{cleveref}
\usepackage{multirow}
\usepackage{capt-of}
\usepackage{siunitx}
\usepackage{graphicx}
\usepackage{gensymb}
\usepackage{slashed}
\usepackage{tabularx}
\usepackage{enumitem}
\usepackage{multirow}
\usepackage{booktabs}
\usepackage{isotope}
\usepackage{xcolor}
\usepackage{bm}
\usepackage{array,mathtools}

\newcolumntype{C}{>{$}c<{$}}
\AtBeginDocument{
\heavyrulewidth=.08em
\lightrulewidth=.05em
\cmidrulewidth=.03em
\belowrulesep=.65ex
\belowbottomsep=0pt
\aboverulesep=.4ex
\abovetopsep=0pt
\cmidrulesep=\doublerulesep
\cmidrulekern=.5em
\defaultaddspace=.5em
}

\definecolor{myred}{rgb}{1., 0., 0.0014072231026931448}
\definecolor{myorange}{rgb}{1., 0.6411371983743057, 0.}
\definecolor{myyellow}{rgb}{0.9998538461538462, 0.8472174696143701, 0.}
\definecolor{mylime}{rgb}{0.8296461722852855, 0.9416163076923076, 0.}
\definecolor{mygreen}{rgb}{0.14471625664756718, 0.9071012307692308, 0.2631632274752619}
\definecolor{myaqua}{rgb}{0.05406425942348916, 0.8614636923076924, 0.7734102885268764}
\definecolor{mysky}{rgb}{0., 0.5883488108126352, 0.9445016153846155}
\definecolor{myblue}{rgb}{0., 0.34480560463730825, 1.}
\definecolor{mynavy}{rgb}{0.25166566047369227, 0., 0.8886266153846154}
\definecolor{mypurple}{rgb}{0.6303632859199042, 0., 0.7270160000000001}

\newcommand{\prn}[1]{ \left(  #1 \right) }

\newcommand{\ord}[1]{\mathcal{O}\left(#1 \right)}

\newcommand{\magn}[1]{\left| #1 \right|}
\newcommand{\mN}{M_{A,Z}}
\newcommand{\mNprime}{M_{A, Z+1}}
\newcommand{\mNj}{M_{A_j,Z_j}}

\newcommand{\mCCth}{m_{\text{th}}^{\beta}}
\newcommand{\mCCthj}{m_{\text{th},\, j}^{\beta}}

\begin{document}

\title{Direct Detection Signals from Absorption of Fermionic Dark Matter}

\author{Jeff A. Dror}
\affiliation{Theory Group, Lawrence Berkeley National Laboratory, Berkeley, CA 94720, USA}
\affiliation{Berkeley Center for Theoretical Physics, University of California, Berkeley, CA 94720, USA}
\author{Gilly Elor}
\affiliation{Department of Physics, Box 1560, University of Washington, Seattle, WA 98195, U.S.A.}
\author{Robert McGehee}
\affiliation{Berkeley Center for Theoretical Physics, University of California, Berkeley, CA 94720, USA}
\affiliation{Theory Group, Lawrence Berkeley National Laboratory, Berkeley, CA 94720, USA}

\begin{abstract}
We present a new class of direct detection signals; absorption of fermionic dark matter. We enumerate the operators through dimension six which lead to fermionic absorption, study their direct detection prospects, and summarize additional constraints on their suppression scale. Such dark matter is inherently unstable as there is no symmetry which prevents dark matter decays. Nevertheless, we show that fermionic dark matter absorption can be observed in direct detection and neutrino experiments while ensuring consistency with the observed dark matter abundance and required lifetime. For dark matter masses well below the GeV scale, dedicated searches for these signals at current and future experiments can probe orders of magnitude of unexplored parameter space.
 \end{abstract}

\maketitle

The search for dark matter (DM) is rapidly expanding both theoretically and experimentally. Weakly interacting massive particle (WIMP) DM searches have pushed the limit on the WIMP-nucleon cross-section near the neutrino floor for masses around the weak scale~\cite{Tan:2016zwf,Akerib:2016vxi,Aprile:2017iyp}. These null results have sparked a renaissance in DM model building, in search of alternative thermal histories which predict lighter DM~\cite{Griest:1990kh,Pospelov:2007mp,Hochberg:2014dra,Hochberg:2014kqa,Kuflik:2015isi,Carlson:1992fn,Pappadopulo:2016pkp,Farina:2016llk,Dror:2016rxc,Dror:2017gjq}. For masses below the GeV scale, DM which scatters off a target will typically deposit energy below the threshold of the largest direct detection experiments ($ {\cal O} ( {\rm keV} ) $), significantly relaxing the direct constraints. 

To discover these lighter DM candidates, the direct detection program is moving toward detecting smaller energy deposits with novel scattering targets and lower-threshold detectors~\cite{Essig:2011nj,Graham:2012su,Essig:2012yx,Essig:2015cda,Hochberg:2016ntt,Derenzo:2016fse,Essig:2017kqs,Budnik:2017sbu,Cavoto:2017otc,Kurinsky:2019pgb}. Current technology is already sensitive to energy deposits of $\ord{\rm eV}$~\cite{Abramoff:2019dfb} and new proposals could detect energy deposits of $\ord{\rm meV}$~\cite{Hochberg:2015pha,Hochberg:2015fth,Hochberg:2016ajh,Schutz:2016tid,Knapen:2016cue,Hochberg:2017wce,Knapen:2017ekk,Griffin:2018bjn}. As the direct detection program pushes the low-mass frontier, it can also broaden its searches for different signals to increase its impact with little additional cost. 

Particle DM detection strategies can be grouped into two classes: scattering and absorption. Searches for scattering look for a DM particle depositing its kinetic energy onto a target within the detector, typically a nucleus or an electron. In contrast, searches for absorption look for signals in which a DM particle deposits its mass energy. Absorption signals have primarily been considered for bosonic DM candidates 
with studies of fermionic absorption signals limited to induced proton-to-neutron conversion in Super-Kamiokande~\cite{Kile:2009nn} and sterile neutrino DM~\cite{Bezrukov:2006cy,Ando:2010ye,Liao:2010yx,Li:2010vy,deVega:2011xh,Long:2014zva,Campos:2016gjh,Lasserre:2016eot} (see also exothermic DM~\cite{Graham:2010ca} and self-destructing DM~\cite{Grossman:2017qzw} for related signals).

In this {\em Letter}, we systematically study direct detection signals from the absorption of fermionic DM. We describe novel signals and their corresponding lowest-dimension operators; project the sensitivities of ongoing and proposed DM direct detection and neutrino experiments to these signals; and demonstrate the consistency of these signals with the issues of DM stability and abundance.

\newpage

\paragraph*{Signals and operators.}
For simplicity, we take DM ($ \chi $) to be a Dirac fermion charged under lepton number, and impose only Lorentz, $ {\rm SU(3)} _C \times {\rm U(1)} _{ {\rm EM}}$, CP, lepton and baryon number symmetries. Baryon number conservation is necessary to avoid proton decays while lepton number allows the (Dirac) neutrino to remain light. We enumerate operators in the effective theory with the fields $  \left\{ \chi , n , p , e , \nu , F _{\mu\nu} \right\} $, where $F _{\mu\nu} $ is the EM field strength tensor. We do not include other QCD resonances as they have no bearing on direct detection.

Consider first dimension-6 operators of the form, $ \left[\bar{\chi} \Gamma _i  \nu \right] \left[ \bar{\psi} \Gamma _j \psi \right]$, where $ \psi \supset \{n ,p , e, \nu\}  $ and $ \Gamma _i = \left\{ {\mathbf{1}} , \gamma _5 , \gamma _\mu , \gamma _\mu \gamma _5 ,   \sigma _{\mu\nu} \right\}    $ denotes the different possible Lorentz structures of the bilinear. These ``neutral current'' operators generate the first class of new signals we consider; $\overset{\textbf{\fontsize{2pt}{1pt}\selectfont(---)}}{\chi } + {T} \rightarrow \overset{\textbf{\fontsize{2pt}{1pt}\selectfont(---)}}{\nu  } + {T}$, where $T$ is a target nucleus or electron which absorbs a fraction of the DM mass energy.
We will focus on nuclear absorption, where the rates may be coherently enhanced, and postpone the study of electron absorption~\cite{NCelectrons}. 

Next, consider dimension-6 operators of the form, $ \left[\bar{ \chi} \Gamma _i e \right] \left[\bar{n } \Gamma _j  p \right] $. These generate a class of ``charged current'' signals; $  \overset{\textbf{\fontsize{2pt}{1pt}\selectfont(---)}}{\chi }+\isotope[A][Z]{X}  \rightarrow e ^{ \pm }+\isotope[A][Z\mp1]{X}_*^\mp $, in which DM can induce $\beta^\pm$ decay in nuclei which are stable in vacuum. 
This process potentially has multiple correlated signals: a detectable $ e ^\pm $, a nuclear recoil, a prompt $ \gamma $ decay from the excited final nucleus, and further nuclear decays if the final nucleus is unstable. 
Induced $ \beta ^+ $ decays have significantly smaller rates relative to $ \beta ^- $ due to the Coulomb repulsion between the emitted $ e ^+ $ and the nucleus, so we focus on DM-induced $ \beta ^- $ decays and leave the $ \beta ^+ $ decays for future work~\cite{Dror:2019dib}~\footnote{$ \beta ^+ $ decays induced by DM with $ m _\chi \gg ~{\rm MeV}  $ were proposed for Hydrogen targets in Super-Kamiokande~\cite{Kile:2009nn}.}. The same charged current operators can also shift the endpoint of the $ \beta^\pm $ distribution for nuclei which already undergo $\beta ^\pm $ decays in vacuum. While this might be detectable at PTOLEMY~\cite{Betts:2013uya,Baracchini:2018wwj}, these kind of experiments have small exposures and large backgrounds, so we defer their study~\cite{Dror:2019dib}. 

Finally, DM candidates which have fermionic absorption signals decay. At dimension-5, the operator $\bar{\chi}   \sigma ^{\mu\nu}  \nu F _{\mu\nu}$ induces decays of $ \chi $ as do the dimension-6 operators, $\bar{\chi}   \gamma ^\nu  \Gamma _{ (5) } \partial ^\mu    \nu F _{\mu\nu}   $, where $ \Gamma _{ (5) } \equiv \left\{ {\mathbf{1}} , \gamma _5 \right\} $. At higher dimensions, there exist operators allowing multiphoton decays. The single photon channel can be detected with the usual line search, while the multiphoton channels are constrained by diffuse photon emission. Detectable fermionic absorption signals, consistent with indirect detection bounds, typically require lighter dark matter as the decay rates scale with a large power of $ m _\chi $. We include a discussion of decays below for each signal and operator we consider. 

\paragraph*{Neutral current signals: nuclear recoils}
We first study the process $\rm \chi + N\to \nu + N$, where N is a target nucleus. We will focus on two operators:
\begin{equation} 
\label{eq:NCops}
\mathcal{O}_{\text{NC}} \, = \, \frac{1}{\Lambda^2}\bar{\chi} \gamma _\mu P _R \nu \prn{ \bar{n} \gamma ^\mu n + \bar{p} \gamma ^\mu p}  + \text{h.c.}\,.
\end{equation} 
We choose this Lorentz structure for concreteness. However, the neutral current signal is not highly dependent on it as long as there is some amount of vector coupling to the nucleons.
These can arise from a theory of a heavy $ Z ' $ coupled to quarks and $ \chi $ with some mixing between the right handed components of $ \chi $ and $ \nu $.
The incoming $ \chi $  is non-relativistic, so its mass dominates its energy resulting in a momentum transfer ($ q $) and nuclear recoil energy ($ E _R $):
\begin{equation} 
\label{eq:NCkin}
q \simeq  m_\chi\,, \quad E_R  \simeq  \frac{m_\chi^2}{2 M}\,,
\end{equation} 
where $M$ is the mass of the nucleus. For contrast, elastic scattering off a nucleus yields at most
$E_R=2 v^2 \mu^2/M,$ where $\mu$ is the reduced mass and $v$ is the DM velocity (see~\cite{Lin:2019uvt} for a recent review). This $1/v^2$ increase in $E_R$ relative to WIMP scattering allows searches for lighter DM with both direct detection experiments and higher-threshold, neutrino experiments.

\begin{figure*}[t!]
\includegraphics[width =\columnwidth]{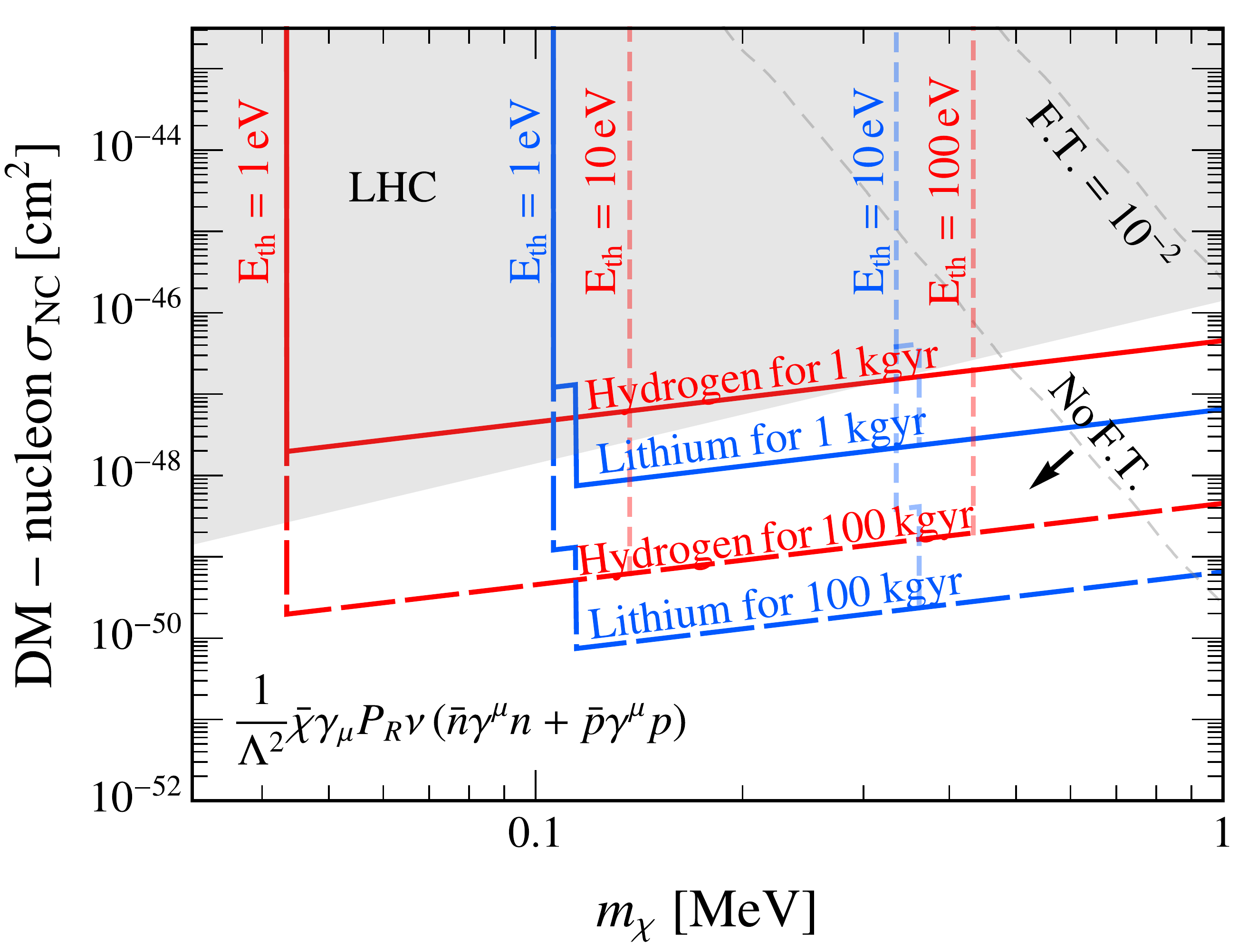}
\includegraphics[width =\columnwidth]{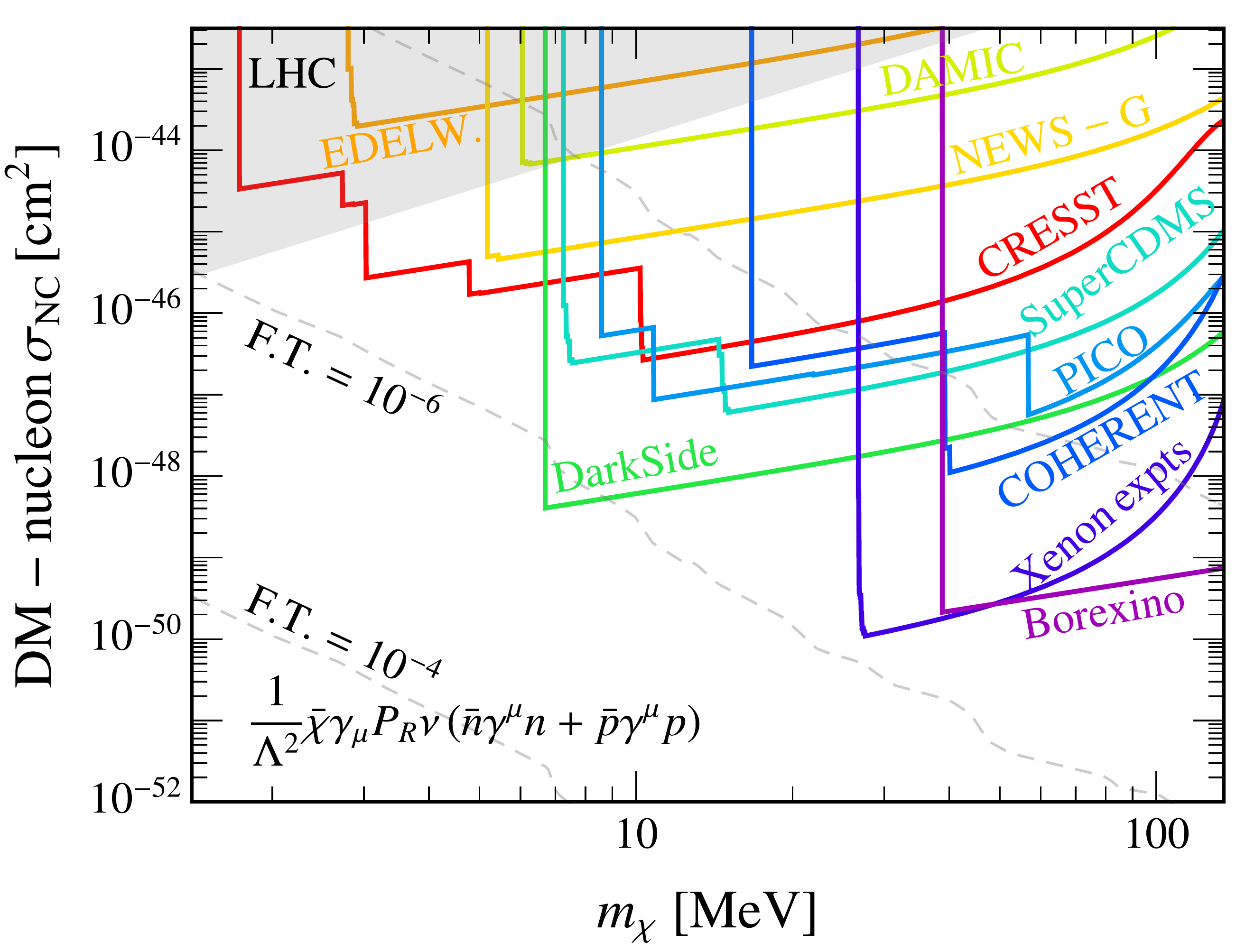}
\caption{\label{fig:NCsigma} 
{\bf Left:} Projected sensitivities of future experiments to $\sigma_{\text{NC}}$. We show two exposures (1/100 kgyr) of two different target materials (Hydrogen in \textcolor{myred}{red} and Lithium in \textcolor{myblue}{blue}) with three possible nuclear-recoil energy thresholds (1, 10, and 100 eV). {\bf Right:} Projected sensitivities of current experiments to $\sigma_{\text{NC}}$, including CRESST III~\cite{Petricca:2017zdp} and CRESST II~\cite{Angloher:2015ewa} (``CRESST'' in \textcolor{myred}{red}); EDELWEISS-SURF~\cite{Armengaud:2019kfj} (\textcolor{myorange}{orange}); NEWS-G~\cite{Arnaud:2017bjh} (\textcolor{myyellow}{yellow}); DAMIC~\cite{Aguilar-Arevalo:2016ndq} (\textcolor{mylime}{lime}); DarkSide-50~\cite{Agnes:2014bvk,Agnes:2018ves} (\textcolor{mygreen}{green}); CDMSliteR2~\cite{Agnese:2015nto} and SuperCDMS~\cite{Agnese:2014aze} (``SuperCDMS'' in \textcolor{myaqua}{aqua}); PICO-60 run with $\rm C_3F_8$~\cite{Amole:2017dex} and PICO-60 run with $\rm CF_3I$~\cite{Amole:2015pla} (``PICO'' in \textcolor{mysky}{sky blue}); COHERENT~\cite{Akimov:2017ade,Scholberg:2018vwg} (\textcolor{myblue}{blue}); LUX~\cite{Akerib:2016vxi}, PandaX-II~\cite{Cui:2017nnn}, and XENON1T~\cite{Aprile:2018dbl} (``Xenon expts'' in \textcolor{mynavy}{navy blue}); and Borexino~\cite{Agostini:2018fnx} (\textcolor{mypurple}{purple}; see \cite{Tretyak:2009sr} to extract nuclear recoil threshold). Both panels include LHC bounds~\cite{Belyaev:2018pqr} and the indirect detection constraints from $ \chi $ decay~\cite{Essig:2013goa} which require different levels of fine-tuning between the UV and IR contributions to kinetic mixing between the photon and $Z'$ for the $ Z ' $ model as described in the text.} 
\end{figure*}

The differential rate of neutral current nuclear recoils from absorbing fermionic DM is; 
\begin{align} 
 \frac{ d R }{ d E _R }&  =N_T \frac{\rho_\chi}{m_\chi} \frac{\overline{ \left| {\cal M} _N \right| ^2}}{16 \pi M^2} \delta ( E _R - E _R ^0 )  \Theta ( E _R ^0 - E _{ {\rm th}} )\,, 
\end{align} 
where $N_T$ is the number of target nuclei, $\rm \rho_\chi \simeq 0.4 \, GeV/cm^3$ is the local DM energy density, $E _R ^0  \equiv m _\chi ^2 / 2M$,  $ E _{ {\rm th}} $ is the experiment's threshold, and $\overline{ \left| {\cal M} _N \right| ^2}$ is the matrix element squared (at $q$) averaged over initial spins and summed over final spins. In elastic scattering, the spread in incoming DM velocities causes a spread in recoil energies, but in fermionic absorption, the rate is sharply peaked at $ E _R = E _R ^0 $. Every isotope in an experiment has a distinct peak with width ($ \Delta E _R $) determined by higher order corrections to Eq.~\eqref{eq:NCkin}, corresponding to $ \Delta E _R  / E _R \sim 10 ^{ - 3} $. There are no modulation signals or rate uncertainties arising from the DM velocity distribution.

The total rate for absorption by multiple nuclei is
\begin{align} 
\label{eq:rate}
R = \frac{\rho_\chi}{m_\chi} \sigma_{\text{NC}} \sum_j N_{T,j} A_j^2 F_j^2  \Theta ( E _{R, j} ^0 - E _{ {\rm th}} ), 
\end{align} 
where $N_{T,j}$, $A_j$, $E _{R, j} ^0$, and $F_j$, are the number, mass number, recoil energy, and Helm form factor~\cite{Lewin:1995rx} (evaluated at $q=m_\chi$ and normalized to 1)  of target isotope $j$. The cross section per-nucleon is $\sigma_{\text{NC}}=m_\chi^2/\prn{4\pi \Lambda^4}$.
Absorption has the unique signature of correlated, peaked counts in $dR/dE_R$ bins containing $E _{R, j} ^0=m_\chi^2/\prn{2 M_j}$ for the different target isotopes with masses $M_j$. This can be a powerful discriminator from backgrounds since the relative heights and spacing of the peaks is completely determined. Whether an experiment can resolve these distinct peaks depends on its energy resolution and the mass splitting between the target isotopes. 

For $ m _\chi \lesssim ~{\rm MeV} $, future experiments are necessary to probe the small nuclear recoil energy. Detailed projections are challenging due to the breadth of proposals and possible absorption by collective modes of nuclei. So, we roughly estimate the sensitivity of such future detectors in Fig.~\ref{fig:NCsigma} ({\bf Left}) where, for simplicity, we require at least 10 events to set our projections, independent of mass or experiment. The cross sections are smaller than those in typical WIMP searches due to the larger number densities of lighter DM. We consider Hydrogen and Lithium targets with energy thresholds of $ {\rm eV} - 100 \, {\rm eV} $ for 1 kg-year and 100 kg-year exposures (see~\cite{Budnik:2017sbu} for one possible realization). In the Lithium target, the detection of two correlated signals from both isotopes is possible. For $ m _\chi \gtrsim ~{\rm MeV} $, we project the sensitivity of current experiments in Fig.~\ref{fig:NCsigma} ({\bf Right}). Interestingly, which experiments best probe the neutral current absorption signal are not always the same as those which best probe WIMPs (e.g., Borexino). Each experiment can only detect the absorption signal off a target isotope when its distinct nuclear recoil energy is larger than the threshold energy, hence the edges in Fig.~\ref{fig:NCsigma}.

We now address the stability of $\chi$. For concreteness, we consider a model where a heavy $ Z ' $ couples in an isospin-invariant way to quarks, with gauge coupling $g_{Z'}$, and to $\chi$ and $P _R \nu$ in an off-diagonal way~\footnote{This suppresses the $\chi \to \nu \nu \nu$ decay; this off-diagonal coupling appears in inelastic dark matter models~\cite{TuckerSmith:2001hy}.}.  Quark loops induce a kinetic mixing, $\epsilon$, between the $ Z ' $ and the photon of order $ \epsilon \sim g _{ Z ' } e / 16 \pi ^2 $ allowing the decay $ \chi \rightarrow \nu e ^+ e ^- $. Without additional $ Z $ or $ Z ' $ mass suppressions, the decay $ \chi \rightarrow \nu \gamma $ is forbidden by gauge invariance while $ \chi \rightarrow \nu \gamma \gamma $ is forbidden as a consequence of charge conjugation (also known as Furry's theorem). For $ m _\chi  \lesssim ~{\rm MeV} $, the electron channel is kinematically forbidden and the dominant decay is $ \chi \rightarrow \nu \gamma \gamma \gamma $, whose primary contribution proceeds through a kinetic mixing and the Euler-Heisenberg Lagrangian, yielding the approximate rate 
\begin{align} 
\label{eq:NCdecaysThreeGamma}
\Gamma _{ \chi \rightarrow \nu \gamma \gamma \gamma } & \,\, \simeq \,\,  \prn{g _{ Z ' } \epsilon}^2  10 ^{ - 19}  \frac{m _\chi ^{ 13}}{ m _e ^8 m _{ Z ' } ^4 } \,.
\end{align} 
Estimating the DM decay rates in this simple UV completion, we find future experiments can quickly probe new parameter space while cross-sections accessible to current experiment are ruled out by indirect detection bounds~\cite{Essig:2013goa}. 

However, it is possible to suppress DM decays by fine-tuning the UV contribution to the kinetic mixing against the IR piece estimated here. Concretely, we define this fine-tuning as $\text{F.T.} \equiv \magn{\epsilon_{\text{UV}}-\epsilon}/\epsilon$ and we show the fine-tuning necessary to evade indirect detection constraints with dashed gray lines labeled ``F.T.'' in Fig.~\ref{fig:NCsigma}. We note that the projected direct detection sensitivities in Fig.~\ref{fig:NCsigma} are insensitive to the details of the UV completion. We study ways to reduce fine-tuning by incorporating flavor-dependent couplings to suppress $\epsilon$ in future work~\cite{Dror:2019dib}.

Also shown in Fig.~\ref{fig:NCsigma} are direct constraints from LHC mono-jet searches on the $ Z ' $ model, which bound new neutral currents below the TeV scale \cite{Belyaev:2018pqr}. Dijet constraints are model dependent: in the $ Z ' $ model, dijet bounds are suppressed in the limit where the quark coupling to the $Z'$ is much smaller than the $\chi$ coupling to $Z'$. The excess neutrino flux emanating from the Sun and Earth due to Eq.~\eqref{eq:NCops} is not quite large enough to be seen in neutrino observatories in the near future. Cosmological bounds depend on initial conditions (e.g., the reheat temperature) and the UV completion. While we postpone a detailed study of the dark matter relic abundance, we comment that a simple way to populate such light dark matter is through its thermal production and relativistic decoupling followed by its dilution from the decay of another heavy particle---a mechanism considered for sterile neutrinos~\cite{Asaka:2006ek}. If the dark matter $\chi$ decouples while it is relativistic, which with only the quark coupling, will occur at the latest around the QCD phase transition, it will be overproduced. After relativistic decoupling of $\chi$ another state becomes non-relativistic leading it to quickly dominate the energy density of the universe. This mechanism can produce and sufficiently dilute DM to achieve its observed relic abundance over the entire DM mass range shown in Fig.~\ref{fig:NCsigma}. While other production mechanisms are possible, one must carefully avoid spoiling Big Bang Nucleosynthesis (BBN) for DM lighter than $\mathcal{O}(1 \text{ MeV})$~\cite{Sabti:2019mhn}, \emph{e.g.} via freeze-in. Neutron absorption of $\chi$ through $n + \chi  \rightarrow p + e $ could also spoil BBN, but the number density of $\chi$ is substantially less than that of SM neutrinos and the rate of scattering is suppressed relative to that of neutrinos by $\prn{m_W / \Lambda}^4 $ making this negligible. 

\paragraph*{Charged current signals: DM-induced $\beta^-$ decays}
Next, consider signals from $ \chi+\isotope[A][Z]{X}  \rightarrow e ^{ - }+\isotope[A][Z+1]{X}_*^{+}$ (or at the nucleon level, $ \chi + n \rightarrow p + e ^- $), which we refer to as an induced $ \beta ^- $ decay. This process can cause stable elements to become unstable in the presence of DM if $ m _\chi $ is large enough to overcome the kinematic barrier. 
Such a signal may proceed through the dimension-6 charged current vector operator, 
\begin{align}
\label{eq:CCop}
\mathcal{O}_{\text{CC}} = \frac{1}{\Lambda^2}\left[  \bar{ \chi} \gamma^\mu e \right]  \left[    \bar{n } \gamma _\mu  p \right] \, + \text{h.c.}  \, .
\end{align}
This can be generated by a $ W ^{\prime} $ which can appear if the electroweak gauge group is embedded in a larger gauge group which subsequently breaks into the SM. 

We consider the vector operator in Eq.~\eqref{eq:CCop} to leverage known results from the neutrino and nuclear physics literature.  The vector-vector interaction primarily induces Fermi transitions which are characterized by their conservation of spin ($ J $) and parity ($ P $) of the nucleus \cite{PhysRevC.99.014320}, also known as $ J  ^P \rightarrow J ^P $ transitions. However, we emphasize that the DM induced $\beta^-$ decay signal is more general, with different vertex structures allowing different transitions. We leave a study of additional interactions to~\cite{Dror:2019dib}.

Denoting the mass of a nucleus of mass number $ A $ and atomic number, $ Z $, by $\mN$, we focus on isotopes which satisfy $\mN < \mNprime^{(*)} + m _e $, such that the nucleus is stable against $\beta^-$ decay in a vacuum (the $ {}^{ ( \ast ) }$ is included to emphasize the daughter nucleus may be in an excited state, typically $200 \,\text{keV} - 1 \, \text{MeV}$ heavier in mass). 
Then DM induced $\beta^-$ decay is kinematically allowed if 
\begin{equation} 
m_\chi > \mCCth \equiv \mNprime^{(*)} +m _e -\mN \,.
\end{equation} 
In these induced decays, $ \chi $ is absorbed by the target nuclei and transfers the majority of its rest mass to the outgoing electron. 
In the limit where $m_\chi  - \mCCth \gg m_e$, the electron and nuclear recoil energies are analogues to the neutral current case with $m_\chi \rightarrow m_\chi - \mCCth $, and are given by;
\begin{align}
\label{eq:CCkin}
&E_{R }  \simeq \left\{ \begin{array}{lr}  m_\chi - \mCCth & {\rm (electron)} \\   \big( m_\chi - \mCCth \big)^2/2 \mNprime^{(*)} & {\rm (nucleus)} \end{array} \right.\,.
\end{align}
Therefore, the energetic outgoing electron will shower in the detector, and can be searched for.  The nuclear recoil energy,  
as with the neutral current case, is 
independent of DM velocity, and can be searched for as well. Additional correlated signals result from the possible de-excitation of the daughter nucleus and its subsequent decay (typically many days later). These multiple signals make possible correlated searches to reduce backgrounds. The specific signals depend on the experiment, the particular isotope, and the DM mass.

\begin{figure} 
  \begin{center} 
\includegraphics[width=\columnwidth]{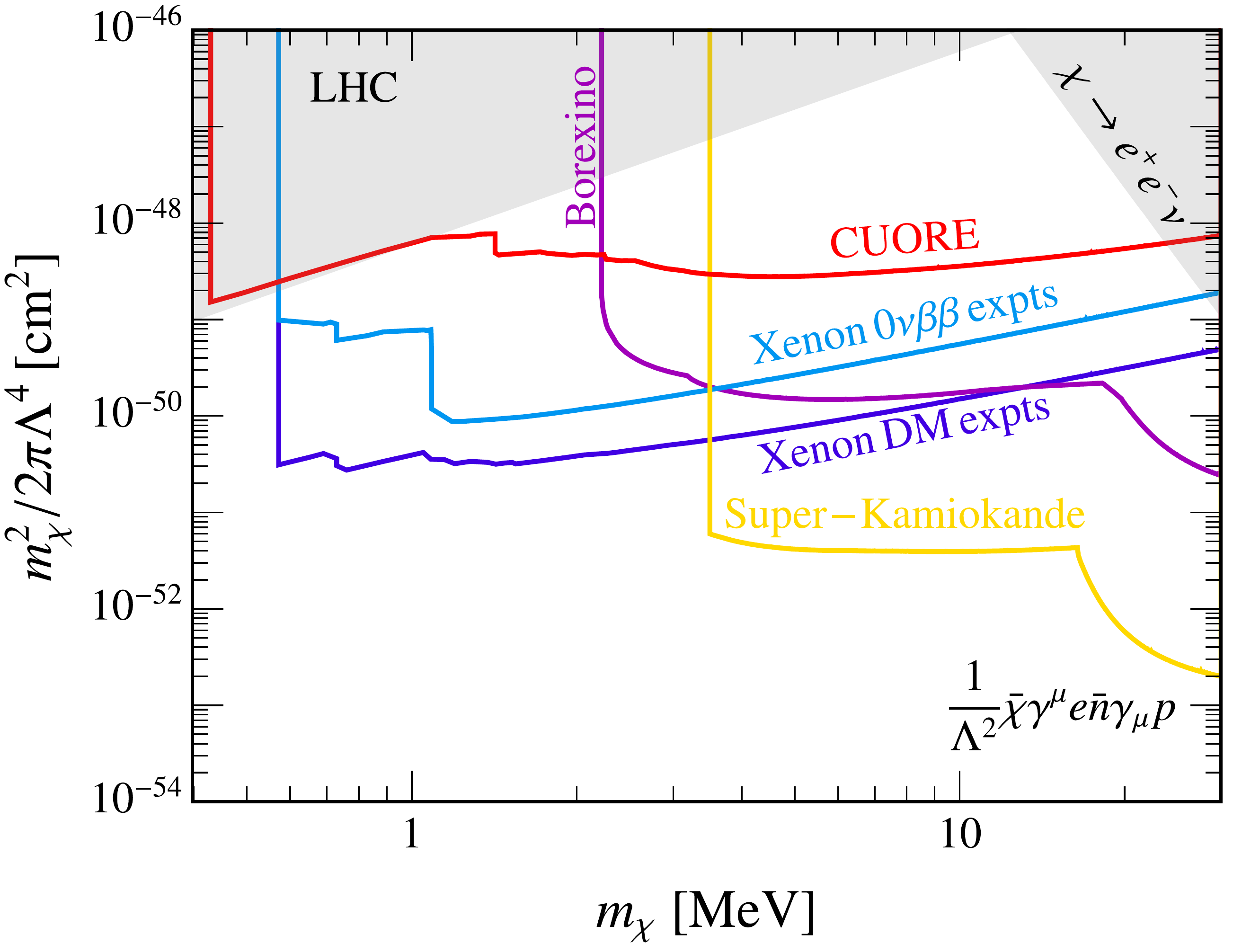} 
\end{center}
\caption{Projected sensitivities to $m_\chi^2/ 2 \pi \Lambda^4$ from a dedicated search for the charged current induced $\beta^-$ transition at
Cuore~\cite{Alduino:2017ehq} (\textcolor{myred}{red}); LUX~\cite{Akerib:2016vxi}, PandaX-II~\cite{Cui:2017nnn}, and XENON1T~\cite{Aprile:2018dbl} (``Xenon DM expts'' in \textcolor{mynavy}{navy blue}); EXO-200~\cite{Albert:2017qto} and KamLAND-Zen~\cite{Shirai:2017jyz} (``Xenon $0\nu \beta \beta$ expts'' in \textcolor{mysky}{sky blue}); SuperKamiokande~\cite{Wan:2019xnl} (\textcolor{myyellow}{yellow}); CDMS-II~\cite{Ahmed:2009zw} (\textcolor{myaqua}{aqua});  DarkSide-50~\cite{Agnes:2014bvk,Agnes:2018ves} (\textcolor{mygreen}{green}); and Borexino~\cite{Agostini:2018fnx} (\textcolor{mypurple}{purple}). Also shown are LHC bounds~\cite{Khachatryan:2014tva} and indirect constraints from $\chi$ decays in our simple UV model~\cite{Essig:2013goa}.
}
\label{fig:Xemaster}
\end{figure}

The rate for DM-induced $ \beta ^{-} $ decays is; 
\begin{align}
\label{eq:CCrate}
R & = \frac{\rho_\chi}{2 m_\chi}  \sum_j N_{T, \, j}   \left(A_j - Z_j \right) \langle \sigma v \rangle_j  \, ,
\end{align}
where we sum over all isotopes in a given target material, $N_{T,\,j}$ is the number of target isotope $j$, and $ \left\langle \sigma v \right\rangle _j $ is an isotope's velocity averaged cross-section
\begin{align}
\label{eq:CCsigma}
 \langle \sigma v \rangle_j = \frac{|\vec{p}_e|_j}{ 16 \pi m_\chi  \,\mNj^2} \overline{  |\mathcal{M} _{N_j} |^2} \, ,
\end{align}
where $|\vec{p}_e |^2_j = (\mCCthj - m_\chi)(\mCCthj - m_\chi - 2 m_e)$ is the electron's outgoing 3-momentum in the center of mass frame (which is approximately the lab frame), in the limit that $m_e, \, m_\chi, \mCCthj \ll \mNj$.
The amplitude $\mathcal{M} _N $ is for absorption by the whole nuclei (the momentum transfer is not enough to resolve individual nucleons), which can be related to the nucleon level amplitude $\mathcal{M}$ (with the spinors normalized to $p _\mu p ^\mu  = \mNj^2$) through the Fermi function, $ {\cal F} ( Z, E _e  )$ and a form factor, $ F _V ( q ^2 ) $:
\begin{align}
\label{eq:CCamp}
\mathcal{M} _N = \sqrt{\mathcal{F}(Z+1, E_e)} F_{V} (q^2) \mathcal{M}  \,.
\end{align}
The Fermi function accounts for the Coulomb attraction of the ejected electron and can enhance the cross-section by several orders of magnitude for heavier elements. The form factor is equal to $  1 $ for small momentum transfer relative to the nucleon mass, $ q ^2 \ll m _n ^2 $, while for larger $ q ^2 $ the dependence can be extracted from the neutrino literature~\cite{Formaggio:2013kya}.
In principle, \eqref{eq:CCsigma} must contain a sum over all possible nuclear spin states. The assumption made here is that this sum will be dominated by $\Delta J^P = 0$ transitions as is the case of a vector coupling \cite{PhysRevC.99.014320}. Excitation of additional final states is possible if $  q  \gtrsim  r _N ^{-1}   $, where $ r _N \simeq 1.2 A ^{1/3} {\rm fm} $ is the nuclear radius~\cite{Engel:1992bf}, however for simplicity we focus on lighter masses such that these do not contribute significantly to the rate for any isotope considered here.

The total rate is found by summing over the contributions from each isotope. Evaluating \eqref{eq:CCsigma} in the limit where $m_e \,, m_\chi \,, \mCCth \ll \mN$, the total rate is;  
\begin{align}
\label{eq:CCrate}
R & = \frac{\rho_\chi}{2 m_\chi}   \sum_j N _{ T, j}   \left(A_j - Z_j \right) \frac{|\vec{p}_e|_j^3 {\cal F} ( Z _j  + 1 , E _e )}{2  \pi \Lambda^4 (m_\chi - \mCCthj )  }   \,,
\end{align}
where we have integrated over all energies with the assumption that such a signal could be detected by most experiments under consideration here given the multitude of correlated high energy signals. 

We project the sensitivity of current experiments to the charged current signal in Fig.~\ref{fig:Xemaster} where we again require at least 10 events to set our projections, independent of isotope mass or experiment. 
Sensitivities are displayed in terms of the theoretically interesting quantity $m_\chi^2/ 2 \pi \Lambda^4$ (to which Eq.~\eqref{eq:CCsigma} reduces in the limit of large $\mNj$ and $m_\chi \gg \mCCthj$, modulo the Fermi function).
As with the neutral current case, limits depend on 
the different isotopes in a given experiment. In particular, the kinks in 
Fig.~\ref{fig:Xemaster} occur at $m_\chi \sim \mCCthj$ for every relevant isotope in a given experiment.

To estimate the DM decay constraints from a typical UV completion, we consider a model with a $ W ' $ coupled vectorially to up and down quarks without any direct couplings to leptons. When kinematically allowed, the dominant decay is $ \chi \rightarrow e ^+ e ^- \nu $ which arises from a kinetic mixing between $ W ' $ and the SM $ W $ boson of order $ \sim g _{W'} e /16 \pi ^2 $. The decay $\chi \to \nu \gamma$ is subdominant since it is at two-loop order, making it roughly $\prn{4 \pi}^2$ smaller. We estimate the decay rate and show the resulting indirect constraints~\cite{Essig:2013goa} in gray in Fig.~\ref{fig:Xemaster}. The decay bounds are much weaker than in the neutral current case as they are suppressed by both the weak scale and the $ W ' $ mass. 

In addition to decays, there are direct bounds from LHC searches for $ p p \rightarrow \ell \nu $. A search was done by CMS at 8 TeV looking for helicity-non-conserving contact interaction models which have contact operators with vertex structure different than that of the SM~\cite{Khachatryan:2014tva} 
which sets a powerful constraint on the charged current operators. For the $ W ' $ model, this constraint corresponds to a scale in Eq.~\eqref{eq:CCop} of $ \Lambda \gtrsim 3.2 ~{\rm TeV} $. In the $W '$ model, there is also a $Z'$ which could lead to direct bounds. However, direct searches for $ Z ' $ are not as stringent as those for the $ W ' $ as the $ Z ' $ can be somewhat heavier than the $ W ' $ and elastic scattering constraints are negligible for the masses and $ \Lambda $ of interest here. We also consider low energy searches for modifications to the $  V-A $ gauge structure of the SM~\cite{Hardy:2014qxa,Gonzalez-Alonso:2018omy} and for light fermions in charged pion decays: $ \pi ^\pm \rightarrow e ^\pm \chi  $~\cite{PIENU:2011aa}, but find they are subdominant to the CMS constraint. Cosmological constraints require detailed assumptions about the initial conditions and the full set of interactions. As for $\mathcal{O}_{\text{NC}}$, a simple production mechanism with $\mathcal{O}_{\text{CC}}$ has the DM decouple from the SM bath while relativistic, followed by late decays of a dominating particle~\cite{Bezrukov:2009th,Patwardhan:2015kga,Evans:2019jcs}. 

\paragraph*{Discussion}
In this {\em Letter}, we have introduced a novel class of signals from fermionic DM absorption in direct detection and neutrino experiments. We have studied the sensitivities of future and current experiments to neutral current signals from the process $\rm \chi + N\to \nu + N$, as shown in Fig.~\ref{fig:NCsigma}. This neutral current causes target isotopes to recoil with distinct energies and correlated rates, enabling significant background reduction in searches. We have also studied the sensitivities of current experiments to induced $\beta^-$ decays from the process $ \chi+\isotope[A][Z]{X}  \rightarrow e ^{ - }+\isotope[A][Z+1]{X}_*^{+}$, as shown in Fig.~\ref{fig:Xemaster}. This charged current enjoys multiple signatures from a sequence of events starting with a nuclear recoil and ejected $ e ^- $, followed by a likely $ \gamma $ decay and often a final $ \beta $ decay or electron capture event several days later. For both signals, ongoing experiments can probe orders of magnitude of unexplored parameter space by performing dedicated searches. 

Without yet knowing the true nature of DM, it is impossible to know how it will appear in an experiment. Perhaps, it has been a fermion, depositing its mass energy into unsuspecting targets all along. 

\begin{acknowledgments}

We thank Artur Ankowski, Carlos Blanco, Tim Cohen, Lawrence Hall, Simon Knapen, Tongyan Lin, Ian Moult, Maxim Pospelov, Harikrishnan Ramani, Tien-Tien Yu, and Zhengkang Zhang for useful discussions, and Vetri Velan, Jason Detwiler, and Lindley Winslow for input on the capabilities of DM direct detection and neutrino experiments. We also thank Yonit Hochberg, Tongyan Lin, Lorenzo Ubaldi, Lindley Winslow, and Tien-Tien Yu for helpful comments on the manuscript. JD is supported in part by the DOE under contract DE-AC02-05CH11231. GE is supported by U.S. Department of Energy, under grant number de-sc0011637. RM was supported by the National Science Foundation Graduate Research Fellowship Program for a portion of this work. 

\end{acknowledgments}

\bibliography{Fabs}

\end{document}